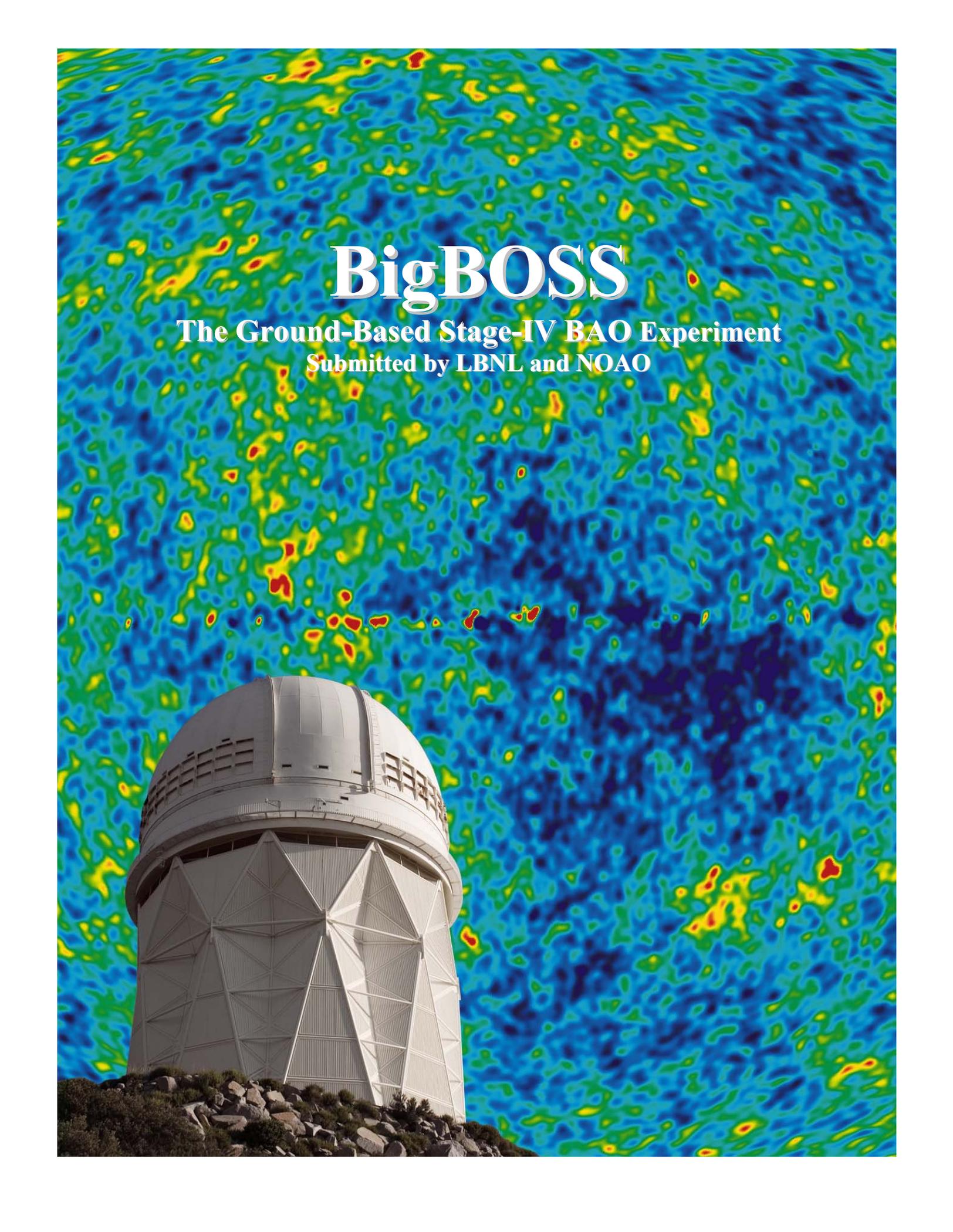

# BigBOSS

## The Ground-Based Stage-IV BAO Experiment

### Submitted by LBNL and NOAO

# BigBOSS: The Ground-Based Stage IV BAO Experiment

This Response to the Decadal Survey
is submitted by:

**The Lawrence Berkeley National Laboratory**
1 Cyclotron Rd MS 50R-5032, Berkeley, CA 94720
David Schlegel, DJSchlegel@lbl.gov, 510-495-2595
Chris Bebek
Henry Heetderks
Shirley Ho
Michael Lampton
Michael Levi
Nick Mostek
Nikhil Padmanabhan
Saul Perlmutter
Natalie Roe
Michael Sholl
George Smoot
Martin White

and

**The National Optical Astronomy Observatory**
950 N. Cherry Ave., Tucson, AZ 85719
Arjun Dey, dey@noao.edu, 520-318-8429
Tony Abraham
Buell Jannuzi
Dick Joyce
Ming Liang
Mike Merrill
Knut Olsen
Samir Salim

# EXECUTIVE SUMMARY

The BigBOSS experiment is a proposed DOE-NSF Stage IV ground-based dark energy experiment to study baryon acoustic oscillations (BAO) and the growth of structure with an all-sky galaxy redshift survey. The project is designed to unlock the mystery of dark energy using existing ground-based facilities operated by NOAO. A new 4000-fiber R=5000 spectrograph covering a 3-degree diameter field will measure BAO and redshift space distortions in the distribution of galaxies and hydrogen gas spanning redshifts from $0.2 < z < 3.5$. The Dark Energy Task Force figure of merit (DETF FoM) for this experiment is expected to be equal to that of a JDEM mission for BAO with the lower risk and cost typical of a ground-based experiment. This project will enable an unprecedented multi-object spectroscopic capability for the U.S. community through an existing NOAO facility. The U.S. community would have access directly to this instrument/telescope combination, as well as access to the legacy archives that will be created by the BAO key project.

The BigBOSS survey will target luminous red galaxies, emission line galaxies, and QSOs. This experiment builds upon the SDSS-III/BOSS project, reusing many aspects of the BOSS spectrograph and computing pipeline designs. The BigBOSS project is enabled by the impressive 3 degree diameter field of view of the 4-m Mayall telescope at KPNO. The focal plane of this telescope will be filled with an automated fiber-positioner capable of targeting 4000 objects simultaneously over a wavelength range from 340 nm to 1130 nm with resolution R=2300–6100. This carefully-designed instrument is capable of measuring redshifts to the brightest [OII] emitters to $z$=2 with a 4-m aperture. Assuming a majority allocation of the dark time and optimal observing conditions during 30% of all nights, and with approximately one-hour exposures, over 5 million targets will be visited per year. We propose to operate for six years at KPNO and then move the instrument to CTIO, the Mayall sister telescope in the southern hemisphere, for a four year run commencing after the Dark Energy Survey (DES) program.

The 30-million galaxy sample of BigBOSS-North provides precision baryon acoustic oscillation measurements over 14000 square degrees from $0.2 < z < 2.0$ and a million QSOs from $1.8 < z < 3.5$. A continuation with BigBOSS-South completes the survey, bringing the total to 50 million galaxies over 24000 square degrees. BigBOSS will sculpt the redshift distribution to maximize the statistical significance of the dark energy measurement. The target selection will be done using existing and planned imaging surveys. A summary of experiment goals is shown in Table 1.

BigBOSS is proposed as a partnership between NSF/NOAO and DOE/OHEP. Details of this partnership will be determined with input from DOE, NSF, and the NOAO user community. Our conceptual plan is that construction would be managed by Lawrence Berkeley National Laboratory, while installation and operation would be managed by NOAO/Kitt Peak. Survey observing support and science operations will be managed jointly.

All of the technology needed for BigBOSS is currently in hand. Construction of the instrument will take three years beginning in 2011. The scope of this project is comparable to that undertaken by DES.

## KEY SCIENCE GOALS

We propose a ground-based survey to measure the expansion rate of the Universe and the growth of structure to sub-percent accuracies between $0 < z < 3.5$. We achieve these constraints by using both the baryon acoustic oscillations and redshift space distortions traced by the large scale galaxy and gas distribution. Such measurements would strongly constrain the phenomenology of dark energy, providing a dark energy FoM ~300 equal to the proposed JDEM mission and substantially exceeding current efforts. A summary of experiment goals is shown in Table 1.

The instrument proposed here for the KPNO 4-m Mayall telescope is capable of simultaneously measuring 4000 redshifts over a 3-degree diameter field of view by employing 4000 individually-actuated fibers feeding spectrographs with wavelength coverage from 340 nm to 1130 nm. The design draws heavily from existing instruments and completed R&D programs, such that no further technology development is required. Over a period of 6 years, we will measure the redshifts of 30 million emission line galaxies in the redshift range $0.2<z<2.0$, covering 14000 deg$^2$. The blue channels of the spectrographs will enable a measurement of the BAO signal in the Ly-$\alpha$ forest for $1.8<z<3.5$ along the lines of sight to a million QSOs. The BigBOSS survey can be extended to the southern hemisphere by moving the instrument to the CTIO 4m Blanco telescope, covering a total of 24000 deg$^2$. Such a survey would be a spectroscopic counterpart to planned wide-field imaging surveys, redressing the current imbalance between imaging and spectroscopy. Such a survey would be transformative in our understanding of the evolution of the galaxy population between $z$~2 and today, much as the Sloan Digital Sky Survey (SDSS) did for the local Universe.

This project enables an unprecedented multi-object spectroscopic capability for the US community, on a telescope platform that resides within the US ground-based OIR system. The US community would have direct access to this instrument and telescope combination, as well as access to the legacy archives that will be created by the BAO key project.

|  | BOSS | BigBOSS-N | JDEM | BigBOSS-N+S |
|---|---|---|---|---|
| Redshift | $0.2<z<0.7$ | $0.2<z<3.5$ | $0.7<z<2.0$ | $0.2<z<3.5$ |
| Sky Coverage | 10000 deg$^2$ | 14000 deg$^2$ | 20000 deg$^2$ | 24000 deg$^2$ |
| Field-of-View | 7.0 deg$^2$ | 7.0 deg$^2$ | 0.6 deg$^2$ | 7.0 deg$^2$ |
| Number of Fibers | 1000 | 4000 | Slitless | 4000 |
| Angular size of Fibers | 2" | 1.5" | n/a | 1.5" |
| Wavelength Range | 360-1000 nm | 340-1130 nm | 1100–2000 nm | 340nm–1130 nm |
| Spectral Resolution | 1600-2600 | 2300-6100 | 200 | 2300-6100 |
| DETF FoM | 57 | 175 | 250 | 286 |
| DETF FoM w/Stage III | 107 | 240 | 313 | 338 |

**Table 1. BigBOSS-North and the full BigBOSS experiment compared to the current BOSS experiment (under construction) and JDEM (the only other stage-IV BAO project currently proposed). The DETF FoMs include Planck priors or Planck plus Stage III supernova and weak lensing experiments per FoMSWG. JDEM (also known as IDECS) FoM entries are for JDEM BAO parameters as determined by the JDEM Science Coordination Group, and are consistent with the version submitted to Astro2010.**

**BAO as a Dark Energy Probe**

The past decade has been one of extraordinary progress in cosmology, with perhaps the most startling discovery being that the expansion of the Universe is accelerating. Such acceleration poses a deep challenge to our understanding of fundamental physics. Distinguishing competing hypotheses for dark energy requires precise measurements of the cosmic expansion history and the growth of structure. The growth of large-scale structure, as revealed in large redshift surveys, has historically been one of our most important cosmological probes. This growth is driven by a competition between gravitational attraction and the expansion of space-time, allowing us to test our model of gravity and the expansion history of the Universe. By using the measurement of baryon acoustic oscillations to map the expansion history and redshift space distortions to constrain the growth history, a large redshift survey can provide sub-percent constraints on dark energy and tests of General Relativity.

Among currently known methods for measuring the expansion history (or distance as a function of redshift) BAO appears to have the lowest level of systematic uncertainty (Albrecht et al. 2006). Sound waves that propagate in the hot plasma of the early Universe imprint a characteristic scale on the clustering of dark matter, galaxies, and intergalactic gas (Peebles & Yu 1970; Sunyaev & Zel'dovich 1970; see Eisenstein, Seo & White 2007 for a pedagogical description). The length of this "standard ruler" can be calculated precisely using physics and cosmological parameters that are well constrained by cosmic microwave background data. Since the acoustic scale is so much larger than the scale of non-linearity (especially at high $z$) the method is highly robust. By measuring this scale with tracers seen at different redshifts we can create a "Hubble diagram" of unrivaled precision. Clustering measurements of the transverse BAO scale yield the angular diameter distance $d_A$ to the measured redshift, while measurements of the line-of-sight BAO scale yield the Hubble parameter $H(z)$. The first clear detection of BAO in the distribution of galaxies was achieved in 2005 by teams from the Sloan Digital Sky Survey (SDSS; Eisenstein et al. 2005) and the Two Degree Field Galaxy Redshift Survey (2dFGRS; Cole et al. 2005). On-going efforts include the Baryon Oscillation Spectroscopic Survey (BOSS; Schlegel et al. 2009), the Hobby-Eberly Dark Energy Experiment (HETDEX; Hill et al. 2004) and the WiggleZ survey (Glazebrook et al. 2007).

**Dark Energy Constraints**

The BigBOSS galaxy sample will allow an exceptional measurement of the acoustic scale and the growth of fluctuations. The dramatically larger volume of BigBOSS, compared to all previous experiments, achieves excellent DETF FoM statistical constraints (Table 1). By using spectroscopic redshifts, BigBOSS leverages the full power of BAO, including the ability to measure the expansion rate, and mitigates potential systematic errors from photometric redshift failures. By measuring the Ly-α forest in a dense grid of bright QSO spectra, BigBOSS will measure the angular diameter distance to, and Hubble parameter at, $z$~2-3, tightly constraining spatial curvature and providing strong tests of inflation and "early dark energy" models. The dense grid of sight-lines will enable measurements that are effectively three-dimensional, leading to a huge gain in statistical power. By mapping large volumes with mildly biased tracers and high sampling density, BigBOSS will provide constraints on the derivative of the linear growth rate that are more than an order of magnitude stronger than we have at present, ushering in an era of precision tests of General Relativity on cosmological scales.

We have calculated, using FoMSWG (Albrecht et al. 2009) matrices and experimental priors, the expected performance of BigBOSS. The DETF figure-of-merit with Planck priors alone is expected to be 286 (176 for just BigBOSS-North); this is to be compared to a JDEM-BAO FoM of 250. If we include Planck plus Stage III supernova and weak lensing experiments as described by FoMSWG, then the FoM is 338 for BigBOSS (240 for BigBOSS-North) as compared to 313 for JDEM-BAO.

BigBOSS comes close to achieving the cosmic variance limit throughout its broad redshift range (see Figure 1a). BigBOSS-South, in particular, will fill-in the $z<1$ regime not covered by BOSS and JDEM. In Figure 1b, we use the FoMSWG principal component analysis to show that BigBOSS will be comparable in performance to JDEM.

In addition to measuring the expansion rate of the Universe, BigBOSS will also measure the rate at which structure forms by mapping the velocity field of galaxies with significantly higher precision and much higher redshift than has been previously possible. This would provide an independent probe of the dark energy, enabling BigBOSS to disentangle whether the acceleration was due to a new field or a breakdown of General Relativity.

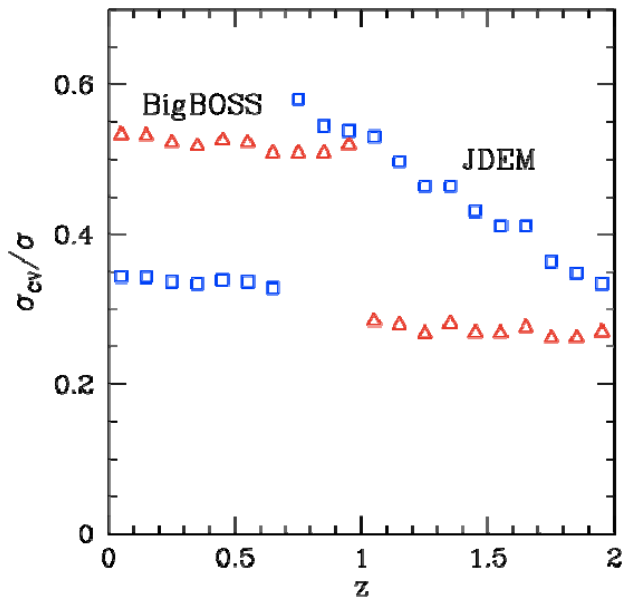
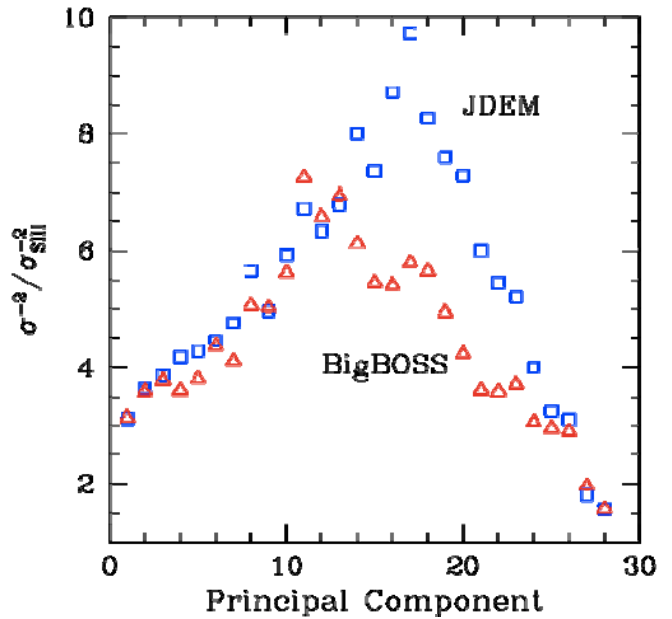

**Figure 1a: Distance accuracies in $\Delta z=0.1$ bins for BigBOSS (red) and JDEM (blue) normalized to the cosmic variance limits. These forecasts were based on the Seo & Eisenstein (2007) Fisher matrix formalism and assume a 50% reconstruction of the acoustic feature.**

**Figure 1b: The inverse variance on the first 30 principal components of the evolution of the dark energy, as defined by the Figure of Merit Science Working Group (FoMSWG). The variances have been normalized to the pre-JDEM Stage III forecasts made by the FoMSWG.**

### Target Selection

The BigBOSS BAO survey targets three classes of objects in progressive redshift ranges – luminous red galaxies for $z<1$, star-forming galaxies for $1<z<2$, and QSOs for $2<z<3.5$.

*Luminous Red Galaxies:* LRGs are a well-studied tracer population of massive dark matter halos, used for BAO measurements in SDSS-I and BOSS. For the BOSS redshift range $0.1<z<0.6$, these are easily selected from SDSS *ugri* photometry. For BigBOSS, the redshift range $0.2<z<1.0$ is easily accessed with the addition of $z$-band photometry. Below $z=0.6$ BigBOSS will target only the LRGs outside the original BOSS footprint. Three years of Pan-STARRS-1 photometry is sufficient to select these objects to the requisite depth with 5% photometric redshift errors. As with BOSS, a sparse sampling of this biased population of objects with a volume density of $3x10^{-4}$ $h^3/Mpc^3$ is sufficient to reach the BAO sample variance limit.

*Emission-Line Galaxies:* The high-resolution channel of BigBOSS is designed to detect either the [OII] doublet in the redshift range $1<z<2$. These emission line fluxes are well-understood to $z$=1.5 from the DEEP2 galaxy survey (Zhu et al, 2008) and the VVDS spectroscopic survey of the COSMOS field (Ilbert et al, 2008). For $z>1.5$, there is agreement of the line fluxes from extrapolation of the [OII] luminosity function and models correlating the rest-frame UV flux (Kennicutt, 1998). The resulting minimal detectable line fluxes for the BigBOSS spectrograph are shown in Figure 2.

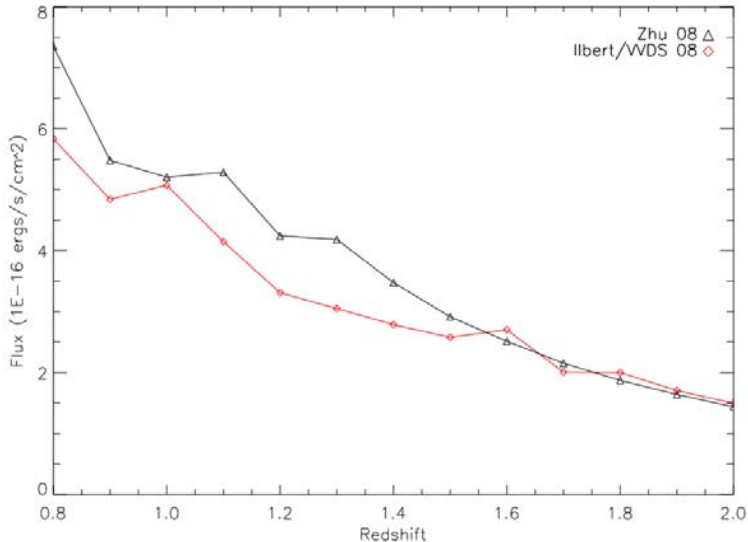

**Figure 2: [O II] line fluxes derived from the DEEP2 luminosity function (black) and the VVDS catalog (red). Both data sources are in agreement over the redshift range 1<z<2, using different extrapolations at z>1.5. BigBOSS is designed for 8-sigma detection of lines at a flux limit of 2.5x10^{-17} ergs s^{-1} cm^{-2} at z=2.**

Efficient target selection of emission-line galaxies is possible from optical imaging surveys. Adelberger et al (2004) showed that the SEDs of these galaxies can be identified with a signature Balmer absorption, and can be separated in color space from late-type galaxies. The color discrimination from the Balmer absorption is accessed through *grz* colors out to $z$=1.5. The expected color cuts and resulting redshift distribution from Pan-STARRS are shown in Figure 3. This color cut achieves >70% efficiency in target selection and an averaged source density d$n$/d$z$ of 2000 deg$^{-2}$.

Target selection in the redshift range $1.5<z<2.0$ is more difficult since the Balmer break is redshifted into the near-infrared. Although it is possible to continue selecting on this feature in *y*-band out to $z$=1.7, NIR measurements become increasingly difficult. Alternatively, the Lyman-break feature of these high-redshift galaxies can be utilized using *ugr* colors out to $z$=2. This will require substantially deeper *u*-band photometry than SDSS.

*QSOs:* The selection of QSOs is straightforward from either color selection or variability. BOSS selects QSOs in the redshift range $2<z<3.5$ with an efficiency of 40% using SDSS colors to *g*=22. This efficiency can be increased with *u*-band data deeper than SDSS. However, variability from Pan-STARRS and other imaging surveys will allow the selection of QSOs with a completeness exceeding 85%.

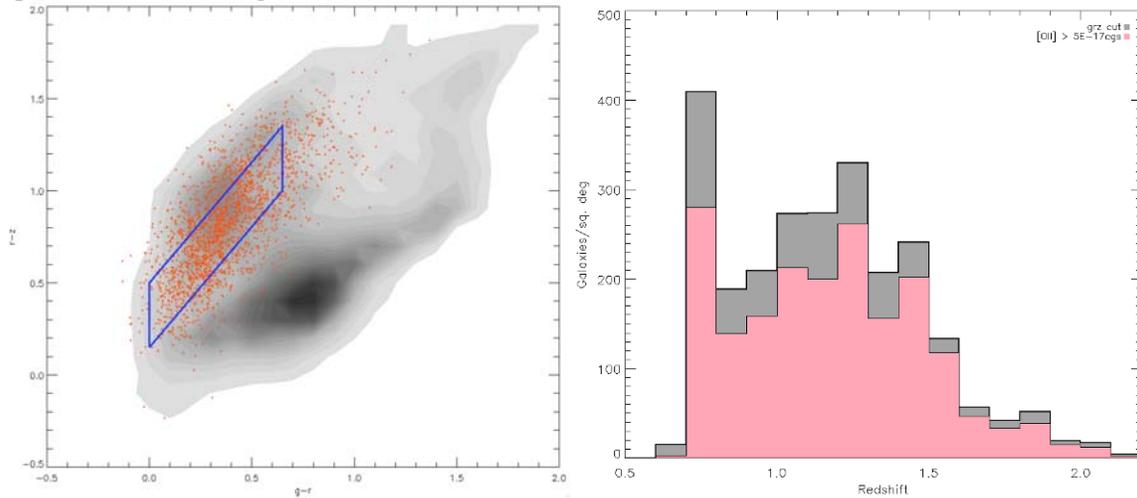

**Figure 3: (left)** *grz* **colors of galaxies using the zCOSMOS fit galaxy SED templates using magnitude errors expected from the PS2 survey. The color cut box selects a sample of strong [OII] emission line galaxies at 1<z<1.5. (right) Surface density of selected galaxies (gray) and the number of galaxies with [OII] flux above the BigBOSS detection threshold.**

## Non-Dark Energy Programs with BigBOSS

The BigBOSS project will have a major scientific impact on US astronomy, both through the products of the core BAO survey, as well as through other spectroscopic studies enabled by the instrument that can be proposed for and carried out by the US astronomical community.

**Cosmology:** BigBOSS provides the largest effective volume yet surveyed for large-scale structure. The galaxy survey will measure 8 times as many Fourier modes as SDSS-I, II and III combined. The BigBOSS Ly-α forest spectra significantly increase both the total line-of-sight distance available for study and the S/N in each spectrum, providing strong constraints on small-scale power. BigBOSS thus provides unprecedented constraints on both large- and small-scale structure, and hence neutrino masses, the running of the spectral index and warm dark matter. The high *z* measurement of $d_A$ and *H(z)* gives us an extremely strong constraint on the geometry of space, and a valuable test of inflationary models which generically predict immeasurably small spatial curvature. A non-zero curvature would inform theoretical ideas about quantum cosmology and the nature of the Big Bang.

**Galaxy evolution:** The BigBOSS primary survey is analogous to the SDSS spectroscopic survey, with the difference that the target galaxy sample will span a larger redshift range at a higher median redshift. As constructed, this survey will represent an unparlleled resource for the study of the evolution of galaxy properties and large scale structure over the last 8 Gyr of the history of the Universe. Based on our current estimates, a one hour exposure time during typical conditions will result in spectra with a signal-to-noise ratio of ~5 (per resolution element between the telluric emission lines) on a *V*=21.5 emission-line galaxy. Hence, these spectra will provide emission line and spectral break measurements for a very large sample of >*L** galaxies at *z*~1.

The survey will enable investigations of the star formation history of the Universe during the last 8 Gyr. This "recent" period in the history of the universe has witnessed a rapid decline in the global star formation density and in the luminosity density emitted by accretion onto black holes. Many mechanisms have been suggested for explaining these perhaps related phenomena, but observational data are currently limited to studies in small volumes (especially at $z > 0.2$). BigBOSS will revolutionize our investigations of galaxy evolution at higher redshift by providing spectra covering the same diagnostic emission lines for millions of galaxies over 1/3 of the sky. In addition, these diagnostics would enable studies of chemical abundances in the star-forming galaxies, the mass metallicity relation, and constraints on the chemical evolution over a large range in redshift and environment. BigBOSS will allow accurate identification of group and cluster environments (a difficult prospect with only photometric redshifts), and can provide constraints on the galaxy merger rate through accurate identification of galaxy pairs.

**QSOs and Ly-α forest:** In addition to the dark energy constraints, the BigBOSS QSO sample will probe the small-scale density fluctuations of the IGM at high redshifts, the ionization background of the Universe, temperature profiles and ionization states of the IGM. A five-fold increase from BOSS QSO survey, BigBOSS will allow precision study of evolution of QSOs by constraining the faint end of the QSO luminosity function at high redshifts. This will provide unparalleled constraints on the lifetimes and hosts of these rare objects and help discriminate among different models of AGN feedback in galaxy formation.

**Galactic Archaeology:** The BigBOSS instrument will be an extraordinary tool to probe the formation and assembly history of the Milky Way. The rich structure revealed by deep imaging surveys of the Milky Way halo (e.g. Newberg et al. 2002; Belokurov et al. 2006; Bell et al. 2008) is widely interpreted to confirm the idea of hierarchical formation of structure in a $\Lambda$CDM cosmology, one of the pillars of modern astronomy. As shown by Johnston et al. (2008), prominent structure seen in coordinate space alone is almost certainly due to very recent satellite accretion, and thus poorly representative of the Milky Way's full accretion history. For a much deeper understanding of the role of accretion in forming the Milky Way, what we need is a large set of data on halo and thick disk stars that includes line-of-sight velocities, Fe abundances, and α-element abundances. This is a task at which BigBOSS will excel.

As demonstrated by the SDSS SEGUE team (Yanny et al. 2009), R~2000 spectroscopy with S/N~30 per resolution element at $\lambda = 8000$Å is sufficient to determine $v_r$ to ~4 km/s and [Fe/H] and [α/Fe] with precision of ~0.2 dex, all of which are excellent for study of the Milky Way halo and thick disk. BigBOSS will be able to take spectra of such high quality for stars with $V=20$ in ~2 hours of exposure. A candidate survey containing ~450 fields would yield ~1.5 million stellar spectra down to $V=20$. For comparison, SEGUE is collecting spectra of 240,000 total stars, a factor of 6 fewer than BigBOSS can collect in ~6 months of observing. SEGUE-2 aims to double SEGUE's outer halo sample, but this will still fall far short of the capabilities of BigBOSS. For identifying accreted satellites in a halo that may be filled with them, large samples are critical. In their simulation of the halo, Sharma & Johnston (in prep.) found that samples of ~100,000 halo stars were needed to identify 5-10% of the accreted satellites. BigBOSS is one of the few proposed instruments capable of readily delivering such samples.

**NOAO PI-led programs:** In addition to the BAO survey, the BigBOSS instrument will lend itself to various smaller scale (i.e., PI-driven) projects. Highly multiplexed spectroscopy will enable studies of galaxy cluster environments, intra-cluster planetary nebulae to map cluster dynamics, the kinematics and binarity in star forming regions within our galaxy, giant stars in the halo of our galaxy and its nearest neighbors, kinematics of open clusters and stellar streams, and constrain bulk flows in the local universe. Spectroscopic monitoring programs with the stable panchromatic spectrographs of BigBOSS can measure stellar masses in binaries, the spatial density variations in the intervening ISM through absorption line monitoring, stellar activity, supernova rates in distant galaxies, AGN reverberation mapping. If coupled with ongoing (e.g., PanSTARRS) or planned (LSST) imaging surveys, BigBOSS will provide invaluable spectroscopic follow-up of targets, especially providing the critical redshift information required for calibrating the photometric redshifts in these surveys. Accurate photometric redshifts are a key component of the weak lensing studies that will provide an independent constraint on dark energy, and of the larger statistical studies of the clustering of faint populations. The resolution of the BigBOSS spectrographs is also sufficient to constrain the evolution of the fine structure constant over the redshift range zero to one.

## BigBOSS in Context

Several other planned and proposed ground-based studies aim to use BAO to constrain dark energy (BOSS, HETDEX, LAMOST, WFMOS), but BigBOSS is in a class of its own (c.f., Table 1). The nearest competition would come from the WFMOS instrument proposed for the Subaru telescope, providing a comparable $A \cdot \Omega$ with an 8-m aperture. BigBOSS is designed for superior BAO performance, with its high throughput in the optical (for $z<1$ LRGs) and R=5000 near-infrared channel to detect [OII] emission line galaxies to $z=2$.

BigBOSS provides a necessary resource for other dark energy experiments. The Dark Energy Survey and LSST will provide independent constraints on dark energy using weak lensing and supernovae. BigBOSS can provide the calibration spectroscopic redshifts to calibrate massive numbers of photo-$z$'s for weak lensing. BigBOSS also has the capability of measuring redshifts for large numbers of supernovae or their host galaxies in its 7 deg$^2$ field

## TECHNICAL OVERVIEW

The BigBOSS concept is to reconfigure the KPNO 4-m Mayall telescope with a new f/5 secondary to deliver a (nearly) 3-degree field of view at a forward cassegrain focus. The focal plane will be populated with 4000 fibers which can be rapidly deployed using a new positioner developed by LBNL. The fibers feed eight 3-armed spectrographs (developed by JHU, and building on the successful heritage of SDSS) that cover a simultaneous wavelength range from 340 nm to 1130 nm at a resolution R=5000. BigBOSS uses existing technology and does not require significant further development; nevertheless, various trades are currently being investigated.

## Telescope and Corrector

The Mayall 4-m telescope at Kitt Peak is well-suited to modification for wide-field fiber spectroscopy across an f/5 3-degree field-of-view. The existing concave hyperbolic primary mirror (M1) would be retained. Instrumentation at the prime focus would be replaced with a convex hyperbolic secondary mirror (M2) designed for wide-field spectroscopy in conjunction with a 3-element corrector/field flattener. The fiber-positioner mechanism locates fiber tips at the focal surface of the telescope. Figure 4 shows the optical layout of the telescope.

The corrector is comprised of two fused silica and one BK7 elements. The first corrector element has an aspheric (4-term) convex front surface, and a concave spherical rear surface. The second corrector element has a concave spherical front (silica) surface, and a concave aspheric (4-term) rear BK7 surface. The interface between the silica and BK7 surface is flat. The average geometric blur is 32 micron (0.3") rms with a maximum of 56 micron (0.5") rms at the field edge. Vignetting (due to the undersized 1.5-m diameter M2) approaches 20% at the edge of the field.

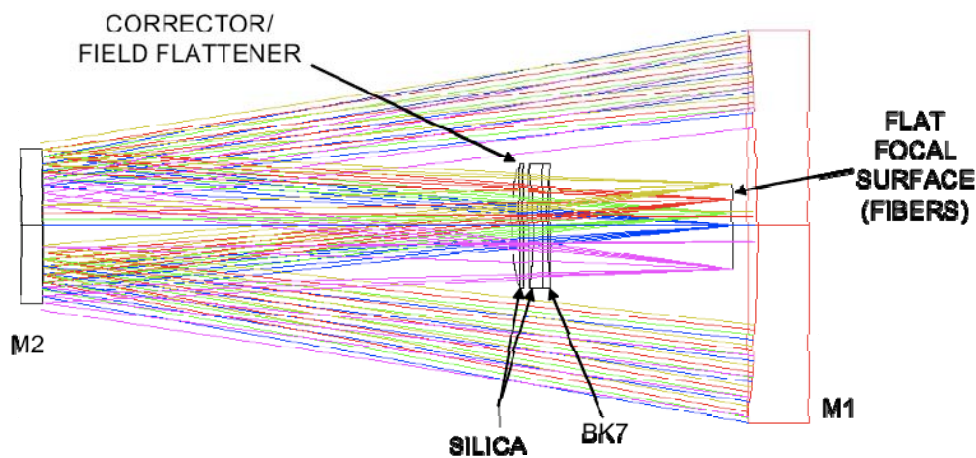

**Figure 4: The telescope is a corrected Ritchey-Chrétien configuration with a 3° field of view.**

## Fiber Positioners and Fibers

BigBOSS makes use of the fiber positioner developed at LBNL during a Laboratory Directed Research & Development program from 2006-2009 (Schlegel & Ghiorso 2008, SPIE 7018-205). The focal plane is defined by an aluminum block machined to the exact shape of the focal surface. The active focal plane is 997 mm to cover the 3-degree field. 4000 circular holes are drilled into this block with a center-to-center separation of 15 mm. Each of the 4000 fiber actuators is plugged into one of these holes from the backside. The astrometric surface is defined by 3 ball bearings (per actuator) on the backside of the block.

Each fiber actuator is mechanically independent of every other. The fiber is moved in a 2-dimensional plane, since the focal surface is very-well approximated as flat within the footprint of one fiber. The motion is controlled by two Micromo 6-mm motors, one driving a rotational axis and the other a translation axis. The fiber motion extends slightly beyond its cell so that all spots on the focal plane can be reached by at least one fiber. The LBNL actuators have achieved the requisite ±20 micron positioning accuracy. The fiber positioning system is shown in Figure 5.

The fibers have 150 micron cores, projecting to 1.5 arcsec diameter on the sky. This choice optimizes S/N for high-redshift, late-type galaxies observed in 1.0 arcsec seeing, or for QSOs observed in the degraded seeing conditions in the blue. The fiber material is Polymicro PBP Broad Spectrum optical fiber, which performs well across the full wavelength range. The maximum attenuation will be 30% at 340 nm in a 15-m fiber run.

## Spectrographs

The spectrographs are a modification of the John Hopkins fiber spectrograph design for SDSS, BOSS, and WFMOS (Smee, Barkhouser, & Glazebrook 2006, SPIE 6269). This is a reverse Schmidt design with a reflective collimator. A dichroic splits the light between two cameras. The BigBOSS design adds a second dichroic, splitting the light between three cameras: blue optimized, visible optimized, and near-IR optimized. Each spectrograph accepts 500 fibers. Therefore, a total of 8 spectrographs are necessary for 4000 fibers.

The BOSS experiment upgraded the SDSS spectrographs with VPH grisms. The use of prisms bonded to the VPH gratings allowed BOSS to re-use existing optical benches. BigBOSS would simplify this design and improve throughput by using VPH gratings without the prisms. The spectrographs will be bench-mounted in the vibration-isolated FTS control room of the Mayall telescope. This will allow significantly improved stability as compared to the Nasmyth-mounted LRIS instrument on Keck, or the telescope-mounted BOSS instrument on the Sloan Telescope.

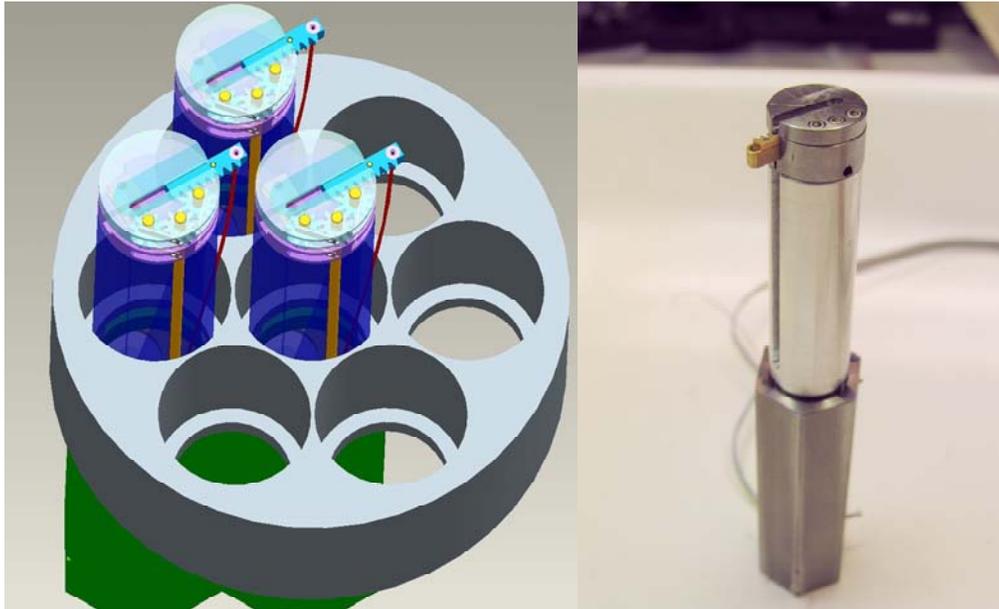

**Figure 5. Image to the left is a CAD drawing showing a few fiber positioners on the focal plane. The image to the right shows a working prototype positioner based upon a servo motor technology. The fiber can be positioned outside of the radius of the positioner itself allowing for full access to any location even outside the circular footprint.**

The red channels cover the wavelength range $800-1130$ nm at resolution 5000, and are used to map emission-line galaxies over the redshift range $0.2 < z < 2.0$ using [OII], [OIII] and Hα lines. The resolution is sufficient to place at least one of the [OII] doublet lines or other available emission lines between bright terrestrial sky lines 96% of the time. The BigBOSS spectral measurement is limited by OH sky continuum. The blue channels cover the wavelength range $340-550$ nm at resolution 3000, and are used to map the Ly-α forest over the redshift range $1.8 < z < 3.5$.

Despite the use of HgCdTe devices in the reddest channel, the instruments are essentially optical spectrographs. The cutoff wavelength of 1130 nm means the optics can operate at room temperature with no significant thermal contribution to the noise. A bandpass filter inside the dewar blocks light at wavelengths beyond 1200 nm.

## Focal Planes

The implementation of the three focal planes per spectrograph builds on components developed under the DOE SNAP R&D programs – CCDs, NIR arrays, electronics, and packaging. The blue focal plane contains a single e2v CCD231-84 $4096^2$ device. These are identical to the BOSS blue-channel devices, with the e2v Astro Broadband coating. At the design dispersion of 0.59 Å per pix, these devices are not the read-noise limited (2 e- demonstrated by BOSS).

| Channel | Detector Format | Dispersion | Wavelength Coverage | Resolution |
|---|---|---|---|---|
| Blue | e2v CCD $4096^2$ 15 μm | 0.590 Å/pix | 340-580 nm | 2300-3800 |
| Visible | LBNL CCD $4096^2$ 15 μm | 0.781 Å/pix | 540-860 nm | 2870-4300 |
| Red | LBNL CCD $2048^2$ 18 μm | 0.732 Å/pix | 820-970 nm | 4590-5360 |
| Red | Teledyne $2048^2$ 18 μm | 0.732 Å/pix | 982-1130 nm | 5360-6120 |
| **Table 2. Detector complement used in each of the three arms of the spectrographs.** | | | | |

The red focal plane contains four 2048x2048 devices arranged in a square. LBNL fully-depleted CCDs cover the wavelength range 820-970 nm, and Teledyne HgCdTe detectors cover the range 982-1130 nm. This split makes optimal use of the two technologies where the quantum efficiencies are highest. This focal plane makes use of the SNAP development, with the detectors mounted at a spacing of 3 mm in a single dewar operating at -140 C.

The readout of the detectors implements the "photons to bits" concept. The electronic modules are co-located with the each detector. A serial digital interface is used for command and control and only a few DC voltages are required. The modules, on command, autonomously digitize data and deliver it over the serial links. The CCD electronics is based on two LBNL developed ASICs, one for processing and digitizing up to four outputs per CCD and the other for clocking and biasing the CCD. The parts are engineered for cryogenic operation and are packaged such that a module plugs directly onto the back of the detector and operates the detector temperature. The module has been demonstrated. The e2v CCD231 is of opposite "polarity" to the LBNL CCD. The appropriate inverting preamp has been built and tested and by design can be seamlessly substituted in a future version of the analog ASIC. The implementation of the e2v clock and bias ASIC is easily accomplished. The NIR electronics is based on the Teledyne SIDECAR ASIC. The SNAP program developed a module that incorporates this ASIC, similar in format to that of the CCD, and supports up to the full 32 channel readout mode. The first parts are in test at this time.

The two types of electronics modules use the same serial communications protocol and physical interface. All the detectors in the three arms can be controlled and readout by one copy of the data acquisition system developed for SNAP. A broadcast convert command generates parallel data streams from the detectors that are buffered and then formatted in a CompactPCI processor. The same processor can configure the detectors and perform local housekeeping chores such a temperature regulation and monitoring. A system simultaneously operating CCD and SIDECAR ASICs has been demonstrated.

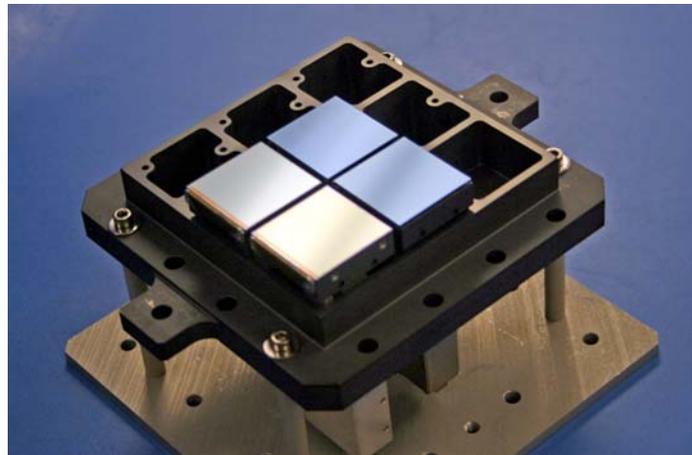

**Figure 6. Shown is a demonstration of the the red arm focal plane built on a SiC cold plate (this prototype was for a 3·3 detector array). The detectors are two LBNL CCDs on SiC mounts (upper right) and two 1.7 μm cut-off HgCdTe NIR detectors, also on SiC mounts (lower left). The protruding boxes at the bottom are the control and digitization electronics for one of CCD and one NIR array.**

The cold plate of each arm is made of SiC. Since all three detector types (LBNL, e2v, and Teledyne) are mounted on SiC (see Figure 6), there is a perfect thermal and mechanical match of materials. A prototype demonstrator was built for the SNAP program. Dimensionally it is very accurate and stable as demonstrated by mechanical vibration studies with detector models in place and thermal measurements with heat injected at appropriate points. Note that in the red arm, the LBNL CCDs and NIR detectors can operate at the same temperature, around 140K, which gives the required low dark currents.

### Survey Strategy

BigBOSS is a dark-time survey that simultaneously observes LRGs, emission-line galaxies and QSOs. Highly-efficient observing strategies are possible with fast reconfigurable fibers. The LBNL fiber positioners in survey mode would split each exposure into sub-exposures, where the tile positioner is moved a fraction of the field size. For example, the telescope would be offset by 1/3 of 3-degree field size in right ascension for each sub-exposure and by $\sqrt{2}/3$ of the field size in declination. Any target is then visible in 9 exposures, and assigned to a different fiber in each of those exposures. A strategy like this can be used to achieve very high completeness, and even recover objects otherwise lost to fiber collisions by alternating between observing each of a pair of close neighbors. The largest efficiency gain may come from allocating exposures to each object only until a redshift is successfully measured. Finally, our experience with SDSS indicates there is no need for extensive calibration exposures for the bench-mounted spectrographs.

### Data Reduction

Data reduction from the BigBOSS spectrograph consists of two largely independent steps: extraction and classification. The extraction step from the raw images results in wavelength-calibrated and flux-calibrated spectra. The BigBOSS data reduction builds directly upon SDSS/BOSS pipelines (idlspec2d and idlutils). These are flexible and modular, employing sophisticated algorithms where necessary, and have proven quite successful for SDSS. Due to the transparency and good documentation of these codes, they have comprised the critical elements of analysis pipelines of other surveys, including the Deep Extragalactic Evolutionary Probe (DEEP) at Keck and Hectospec at the MMT.

## TECHNOLOGY DRIVERS

The BigBOSS conceptual design intentionally makes use of existing technologies. The current technology drivers are (1) enabling a very wide-field survey capability on a 4m aperture; (2) the LBNL fiber positioner technology; (3) the fully-depleted LBNL CCD technology; and (4) the high-throughput spectrographs developed by JHU.

Trade studies may also identify possible cost savings with alternative technology choices. Other considerations may allow somewhat increased wavelength coverage, enabling other science. These trade studies include:

1. The secondary mirror polishes an existing Zerodur blank. A Hextek mirror could be considered as a lower-cost option, although it may well result in unacceptable image quality.
2. Cost comparison of the WFMOS and LBNL fiber positioning designs.
3. Modifications to the spectrograph design could be considered to allow larger beam angles, and therefore larger wavelength coverage at the design resolution. This could increase the science return by measuring Hα to higher redshift.
4. The size of the focal plane presents some challenges to the telescope baffling. A thorough trade study of vignetting may suggest that the field size be reduced somewhat below the current design of 3 degrees. A more thorough trade study may suggest a somewhat smaller field-of-view would be preferred.

## ACTIVITY ORGANIZATION, PARTNERSHIPS, AND CURRENT STATUS

**Partnership**: BigBOSS is a proposed partnership between the NSF (through NOAO/Kitt Peak) and DOE. The dark energy science case has been developed by the DOE Lawrence Berkeley Lab to fulfill the community mandate of a Stage IV BAO experiment. NOAO is involved in developing this science case, as well as ensuring the fair community use of their telescope resources. As such, it is expected that a public data release policy would be implemented, piggy-back programs would be solicited, and some time would be made available through a traditional PI-led program.

**Management Plan:** The BigBOSS project will be jointly managed by the DOE and NOAO, with Project Offices at LBNL and Kitt Peak. The DOE Project Office will have primary responsibility for the instrument construction, including the optics, fiber system, spectrographs and spectroscopic pipeline software. The Kitt Peak Project Office will have primary responsibility for instrument installation, telescope modifications and mountain-top operations. A Director to be selected in consultation between LBNL, NOAO and the funding agencies will provide overall management direction, with responsibility for budget, schedule and formation of the BigBOSS collaboration.

The BigBOSS collaboration will be structured to attract institutions with a strong scientific interest in the data products that can make significant contributions to the project. DES and SDSS-III, both of which are jointly funded by DOE and NSF, provide recent examples of successful collaboration building to carry out large astrophysical surveys. BigBOSS will follow these models to form a strong team for the construction, operations and science. The time scale for BigBOSS is complementary to that for DES and BOSS, and we anticipate that many of the same institutions involved in those projects will be interested in joining BigBOSS. The BigBOSS collaboration will have a scientific oversight committee with responsibility for the target selection, survey strategy and execution, and scientific management.

Community access to the instrument will be accommodated through a competitive selection process managed by NOAO.

**Public Access:** While the main science focus is the BAO survey aimed at constraining the redshift evolution of the dark energy parameters, BigBOSS enables an unprecedented multi-object spectroscopic capability for the US community. We envision that this capability will be available directly for use by the US community, through the normal time allocation process operated by NOAO. In addition, since the BAO key project aims to cover a very significant fraction of the extragalactic sky, many projects could be proposed to "piggy-back" on the BAO survey, either through dedicating some fraction of the fibers, or to change the cadence and pointing of the survey observations. Lastly, since there is significant synergy between various galaxy evolution and large scale structure studies and the BAO survey, as well as significant overlap in the targeting strategies, the BAO surveys can be designed in consultation with community-led teams to define the optimal survey with a broad and rich science yield.

The BAO survey will result in a dataset of significant legacy value. A database consisting of spectroscopic redshifts and emission line measurements of over 50 million galaxies will be the spectroscopic equivalent of the digital sky survey and, in concert with various imaging databases (SDSS, PanSTARRS, LSST, etc.) will be revolutionary for studying galaxy evolution and structure formation between redshifts of 1 and the present. In addition, other survey data resulting from the BigBOSS spectrographs (e.g., surveys of the Milky Way thick disk and halo, M31, clusters of galaxies, etc.) will also be archived and available for public use. Hence, it is envisioned that the US community will obtain significant benefit from this collaboration and be a integral partner in its execution.

**Status**: BigBOSS is currently at a pre-proposal stage. The substantial community interest in this project is expected to lead to a distributed multi-institutional collaboration, modeled after the successful SDSS/BOSS collaboration. Details of this partnership have not been finalized, and will require DOE and NSF review.

## ACTIVITY SCHEDULE

The BigBOSS project is divided into four major phases:

1. An R&D Phase beginning in FY09 and continuing through the first quarter of FY11.
2. A Construction Phase that begins with a proposed CD-0 at the beginning of Q2-FY11 and continues to CD-4 at the end FY14.
3. A Northern Operations Phase at KPNO that includes 6 years of instrument operations.
4. A follow-on Southern Operations Phase at CTIO that includes 4 years of instrument operations.

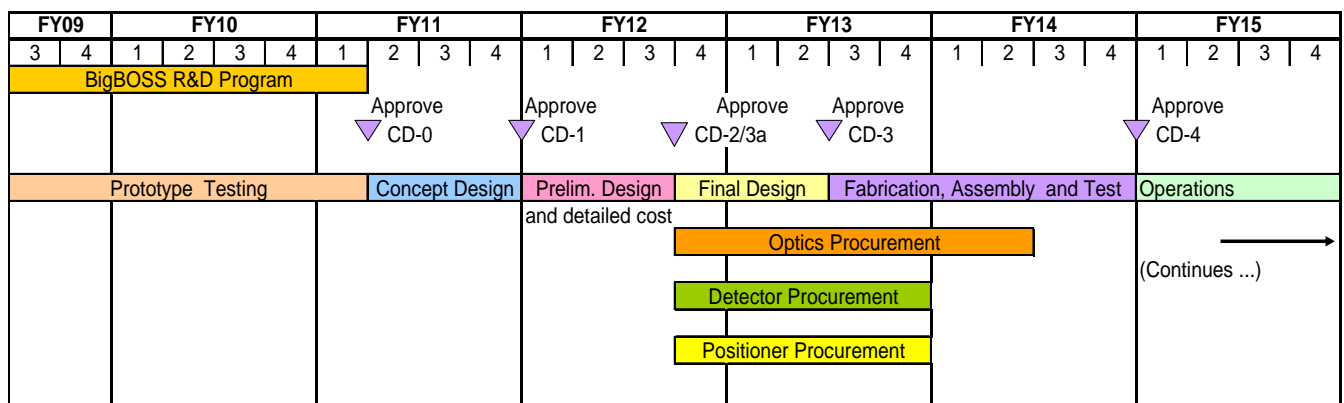

**Figure 7. BigBOSS Top-Level Schedule showing key decision points and major procurements.**

Figure 7 shows the proposed schedule for the BigBOSS project. We start with an initial R&D Program extending through the first quarter of FY11 and funded by the DOE. During this time prototypes of critical portions of the spectrograph will be constructed and evaluated and the existing prototype of the positioner will be refined and additional testing done. Optical designs will also be refined and more detailed costs obtained for all elements. This will put the project in a strong position for a CD-0 at the beginning of FY11 Q2 and allow development of a reliable cost range for a CD-1 at the beginning of FY12 Q1. As a result of prototyping completed in the R&D period we will have a solid preliminary design and a final cost in time for a CD-2 at the beginning of FY12 Q4.

We propose that CD-2 be combined with a CD-3a to start the optics fabrication due to the long lead time required to manufacture the 1.5 meter secondary mirror. Further schedule risk reduction will be achieved by also beginning the detector and positioner procurement at that time. Three more quarters will provide time to complete the final design of all elements to support a CD-3 and start of construction at the beginning of FY13 Q3. An additional 18 months is required to complete the fabrication, assembly, test and installation of all elements to support a CD-4 and start of operations at the beginning of FY15 Q1.

The operations phase consists of a six year period of instrument operations at KPNO. The instrument is then dismounted and shipped to CTIO for installation and a potential four year period of operations there. For the latter, no changes to the optics, secondary or instrument are needed.

## COST ESTIMATE

The total project costs have been estimated using a bottoms-up approach relying on recent quotes and labor estimates from similar systems. These include the BOSS spectrograph upgrade, recent budget estimates for the WFMOS spectrograph, and budget estimates for detectors, electronics and ground support equipment for DES and other LBNL instrument programs. Table 3 below shows the cost break down for the project.

**R&D Phase:** During the approximately 2-year R&D phase, the optics design will be refined and optimized, a prototype of the 3-arm spectrograph will be fabricated and tested, and the existing fiber positioner prototype will undergo further testing and cost optimization. Project costs for the R&D period include non-base funded engineering support along with parts and

| | Table 3 -- BigBOSS Cost Breakdown.* | | |
|---|---|---|---|
| | | Cost Phase ($M FY09) | |
| **WBS #** | **Description** | **R&D** | **Construction** |
| 1.0 | Project Management & System Engineering | | |
| 2.0 | Spectrographs and Instrument Elect. | | |
| 3.0 | Fiber System with Positioners | | |
| 4.0 | Optics | | |
| 5.0 | Data Pipeline | | |
| 6.0 | R&D Phase | | |
| 7.0 | Contingency | | |
| Totals | | | |
| | Grand Total | | |

* Cost data will be available through the NAS/Astro 2010 website.

materials for the construction of the prototypes.

**Project Management**: A full time Project Manager with an administrative assistant and 0.5 FTE of scheduler support, as well as a full time Systems Engineer with of additional engineering support is expected for a project of this scope. Costing of this element was done by assuming a management staff comparable with that of similar sized projects, such as DES and Daya Bay, and using current LBNL labor rates.

**Spectrograph:** We estimated the cost of eight spectrographs and their data processing and control electronics. Costs for the structure and optics bench are JHU estimates based on the design and fabrication of the very similar BOSS spectrographs with updated costings for the WFMOS spectrographs (Smee et al. 2006). The spectrograph optics costs are also based upon the BOSS and WFMOS designs, with an adjustment made to account for three arms on the BigBOSS. Dewar and vacuum system costs are based on very similar systems recently fabricated for BOSS, and similar systems developed by LBNL for other projects. Detector costs are based on:

1) Recent quotes from e2v for the 4kx4k part used in the short wavelength arm.
2) Recent fabrication and packaging costs for the 4kx4k LBNL parts.
3) Recent quotes from Teledyne for 2kx2k 1.7 micron cut-off NIR devices.

The cost for CCD front-end electronics is based on the use of existing CCD readout electronics with minor engineering modifications to accommodate the two types of CCDs. Also included is engineering for parts mounting and assembly testing. Costs for the NIR front-end electronics include engineering for parts packaging and assembly testing as well as procurement of SIDECAR ASICs based on recent quotes from Teledyne.

Cost for the detector interface and spectrograph control electronics is based on a similar electronics system that was developed, fabricated and tested for LBNL current projects. Spectrograph readout and control software is also included in this WBS, modeled on a similar system for BOSS.

**Fiber System:** The fiber system consists of an assembly of the 4000 robotically positioned optical fibers including the positioners, fibers, and electrical control. The cost of the optical fibers is based on an assumed length of 15 meters for all fibers to accommodate the worst case run from a positioner to its corresponding spectrograph, along with the purchase of 12.5% spares. The total cost for the fibers includes labor to attach a ferrule to the positioner end and to construct the linear arrays of 500 fibers which are mated to each spectrograph. Fiber harnesses that are being built for BOSS provided an accurate cost model for this assembly. Cost for the tray support system is based on an LBNL engineering estimate to design and build a system meeting the special requirements on the support of the optical fibers. To reduce cost and schedule uncertainty of the positioner elements, a prototype system was designed, fabricated and extensively tested. Positioner costing is based on invoice costs of the purchased components and machined parts of the prototype assembly.

**Optics:** All instrument-specific optics installed into the telescope structure are grouped here. This includes an upper optics assembly comprised of a new secondary mirror with its associated support structure and fiber position camera mounted above the secondary mirror, and a lower optics assembly comprised of the cassegrain cell assembly, the atmospheric dispersion corrector assembly, and the focal plane plate on which the 4000 fiber positioners are mounted. The guider and auto focus modules mounted on the focal plane are also included. Cost of the secondary mirror is based on a quote from Sagem/REOSC. The proposed vendor has extensive experience with a highly aspheric and convex mirror such as this which presents special challenges for testing and which are reflected in the cost estimate. The fiber positioner camera is a Fairchild product developed around a 9kx9k CCD produced for a military customer (a large format room-temperature camera). Costing for the cassegrain cell assembly and the atmospheric distortion corrector assembly was done by using costs for similarly sized optics element of similar precision developed by KPNO. Cost of the focal plane plate is an LBNL engineering estimate based on the size of the part, the number of machined features required to mount the plate and the 4000 positioners, and the tolerances required.

LBNL engineering estimates, in consultation with KPNO, were also made for the structures for both the upper and the lower optic assemblies and are based on the size of the structure needed to attach the various elements to the existing telescope structure, the number of mounting features required, and the tolerances needed. The guider and auto focus modules are costed assuming a semi-custom design tailored to the requirements on the project and built around a standard CCD. They will use the same front-end electronics modules used for the CCDs on the spectrographs.

**Data Pipeline:** We have estimated the Construction Phase costs for setting up the spectrographic pipeline. It includes pre-operations software infrastructure development cost, but does not include operations cost or other costs post CD-4. Data reduction from the BigBOSS spectrographs consists of two largely independent steps: extraction and classification. These build directly upon the BOSS pipeline currently in development. Pipeline operations costs are included in the Operations budget.

**Operations:** The operations budget includes only the labor cost directly attributable to the instrument operation and the incremental costs incurred at KPNO for the new instrument. KPNO estimates that additional personnel above that required for current telescope operations will be required for support of the BigBOSS instrument. We also estimate maintenance costs for parts and repairs. LBNL technical and computing support of the instrument operation and computing pipeline is estimated to require an additional support. We currently estimate shipping and reinstallation of the instrument at CTIO to be small including some minor mechanical modifications required. The estimates are based in part on the experience of running the KPNO telescopes (WIYN, Mayall, 2.1-m) and the SOAR telescope.

**Relevant Decadal Survey White Papers**

#35  Cahn, Robert, "For a Comprehensive Space-Based Dark Energy Mission"
#70  Eisenstein, Daniel, et al., "Cosmology from a Redshift Survey of 200 Million Galaxies"
#106 Gunn, James E., et al., "Understanding the Astrophysics of Galaxy Evolution: the role of
      spectroscopic surveys in the next decade"
#249 Riess, Adam, et al., "Dark Energy from a Space-Based Platform"
#314 White, Martin, et al., "The Baryon Oscillation Spectroscopic Survey: Precision
      measurement of the absolute cosmic distance scale"